\documentclass[conference]{IEEEtran}

\usepackage{pdfpages}
\usepackage{graphicx}
\usepackage{epsf} 
\usepackage{subfig}
\usepackage{algorithm}
\usepackage{algorithmic}
\usepackage{fancyhdr}

\ifCLASSOPTIONcompsoc
  \usepackage[nocompress]{cite}
\else
  \usepackage{cite}
\fi

\usepackage{enumitem}

\IEEEoverridecommandlockouts

\begin{document}

\title{Blockchain Based Residential Smart Rent\\ \vspace{3mm}}

\author{
\IEEEauthorblockN{Andr\'e S. Proen\c{c}a, Tiago R. Dias, Miguel P. Correia}
\IEEEauthorblockA{INESC-ID, Instituto Superior T\'ecnico, Universidade de Lisboa, Portugal}
\IEEEauthorblockA{Unlockit.io, Portugal}
\IEEEauthorblockA{\{andre.proenca, miguel.p.correia\}@tecnico.ulisboa.pt, tiago.dias@unlockit.io}
}

\maketitle

\thispagestyle{plain}
\pagestyle{plain}

\begin{abstract}
The real estate market includes complex and inefficient mediation processes. Renting a property envolves multiple entities with different responsibilities and interests. Therefore it is imperative to establish a trustful relationship between parties through intermediaries such as notaries, banks or real estate agencies to avoid eventual disputes. Although an intermediary ensures trust, the current process still has some drawbacks concerning efficiency, costs, transparency, bureaucracy and data security. The blockchain technology aims to reduce this issues by providing transparent and secure real estate transactions. We propose a GDPR compliant blockchain-based residential smart rental platform, designed to allow both landlords and tenants to establish rental contracts and make rental payments securely. \textit{Index Terms -- Blockchain, Smart Contracts, Real Estate, GDPR}
\end{abstract}

\section{Introduction}

Blockchain technology in recent years has been growing in popularity at a vertiginous pace, mainly due to the 2008 white paper entitled \emph{Bitcoin: A Peer to Peer Electronic Cash System} \cite{nakamoto2008bitcoin} published pseudonymously by \emph{Satoshi Nakamoto}. This whitepaper proposed an innovative network which became the first ever blockchain application, broadening horizons by unlocking a whole new spectrum of innovative possibilities reshaping some sectors of society, such as the real estate.

The Blockchain technology key features include network decentralization, data persistence, data immutability, and transaction traceability \cite{yaga2019blockchain}. Together, they compose a secure system that solves traditional real estate rental problems such as high intermediation costs, time-consuming bureaucracy, security flaws and fraud.

The traditional establishment of a rental requires an agreement between at least two parties that should be sealed through a contract. This process can be adapted to the blockchain through smart contracts. Despite the technological advancements, regulations in the great majority of nations, have yet to be discussed, making the process of renting real estate assets, through blockchain, highly unpredictable under the law. Nevertheless, the development of applications in the European Union, must comply with the general data protection regulation (GDPR \cite{oficialgdpr}). As a consequence, privacy issues must be taken very seriously when building GDPR-compliant blockchain-based systems.

This paper presents a platform design, conceived for Unlockit, a DLT company operating in the real estate industry. We propose a GDPR-compliant blockchain-based residential smart rental platform, that enables the establishment and management of residential rental contracts with a high level of security and transparency. The platform main features include digitally signing contracts and automating rental payments.

\section{Related Work}

Renting a residential property requires a written contract, agreed upon and signed by two parties, a tenant and a landlord. A contract should outline the terms and conditions of a rental, which includes the obligations and rights of both parties, respecting the laws of the country where the property is located. Contracts can be defined according to duration, such as fixed or short term, and property type, such as the entire property or a single room  \cite{ContractsCounsel}. 

Rental contracts, in general, do not demand the engagement of intermediaries, however, they are frequently validated with the support of real estate agents and lawyers \cite{garcia2020legal}. In the context of residential renting, blockchain aspires to behave like an intermediary, capable of notarising rental agreements by ensuring their validity and certifying that they were agreed upon, on a certain date. It intends to provide a transparent, immutable, traceable, and intermediary-free rental system capable of reducing intermediary costs, bureaucracy, and fraud.

Blockchain also referred to as a distributed ledger is an append-only data structure similar to a distributed database, responsible for maintaining an history of transactions \cite{zheng2017overview}. A transaction is a transfer of tokens, between peers and are usually designated as cryptocurrencies, however, they may represent a way of registering assets such as real estate properties or rental contracts. The blockchain  is maintained by a group of peers, who do not trust each other. Transactions, when validated, are grouped into blocks and cryptographically chained with a continuously growing sequence of blocks. New blocks are inserted by peers who own a copy of the blockchain, and all peers must come to a consensus such as proof-of-work \cite{yaga2019blockchain} or proof-of-stake \cite{saleh2021blockchain} before updating the state of the blockchain. 

Blockchain can be divided in two categories based on permission type, permissionless and permissioned \cite{taskinsoy2019blockchain}. Permissionless such as Bitcoin \cite{nakamoto2008bitcoin}, Ethereum \cite{buterin2016ethereum}, or EOS \cite{grigg2017eos}, are fully transparent, allowing anonymous peers to join and modify the ledger without requiring approval from an authority, whereas, permissioned suchs as Hyperledger Fabric \cite{dhillon2017hyperledger}, R3 Corda \cite{brown2016corda} and Canton \cite{canton20}, allow only a restricted set of identified peers to join, and modify the ledger. 

A smart contract is an executable code that resides and executes on the blockchain to enforce the terms of an agreement \cite{alharby2017blockchain}. It ensures automated enforcement of the terms whenever certain criteria are satisfied \cite{buterin2014next}, and features self-verifying, self-executing, tamper-proof, traceable and irreversible properties \cite{mohanta2018overview}.

Implementating a smart contracts depends on the blockchain platform. 
Hyperledger Fabric is a permissioned blockchain platform designed for business applications that supports pluggable consensus protocols. It offers high levels of confidentiality, resilience, flexibility and scalability. Its network consists of nodes whose identities are given by a membership service provider. There are three different types of nodes that undertake different responsibilities, namely client nodes, peers nodes, and ordering service nodes. The peer nodes maintain the ledger state which is used by the client nodes to propose transactions to execute and broadcast them for ordering. The ordering service nodes set the order of all transactions. However, they do not interact in the execution or validation processes \cite{polge2021permissioned}. The Hyperledger Fabric ledger consists of two merged components, the world state, maintaining the latest transaction data, and the transaction log, maintaining the transaction history resulting in the world state \cite{Fabric}. 

Crypto wallets are a new type of digital wallet, offering a secure environment for accessing and transacting on blockchains \cite{houy2023security}. They store private keys or seed phrases, essential for authorizing and signing transactions \cite{jorgensen2022universal}. 

GDPR or General Data Protection Regulation \cite{oficialgdpr} is an EU data privacy and security regulation, that sets standards for collecting and processing personal data of EU residents that contract services or goods. Therefore, a company providing services to EU residents, may face high monetary penalties if fails to process consumer information accordingly \cite{gdpr}. The applicability of GDPR to blockchain raises questions regarding personal data, since blockchain provide transparent and immutable records \cite{gdprAndBlockchain}. Blockchain-based companies may comply with GDPR by employing a permissioned blockchain along with off-chain storage mechanisms.

A smart rental contract is regarded as the latest paradigm for renting a property, based on the blockchain technology. Existing or developing projects such as Rentible \cite{rentible}, Smart Realty \cite{SMARTRealty}, Midasium \cite{midasium}, RentPeacefully \cite{rentpeacefully} and the The Bee Token \cite{beetoken} are implementing smart rental platforms through different approaches. Although innovative, they lack core properties such as transaction privacy and compliance with the EU GDPR regulation.

\section{Platform Design}

The current real estate rental process has several issues that our proposed platform addresses, namely: 

\begin{itemize}
    \item High intermediation costs
    \item Time-consuming and over-bureaucratic processes
    \item Information shared over multiple unsecured channels
    \item Lack of transparency and traceability
    \item Lack of automatic rental payment mechanisms
\end{itemize}

\subsection{Overview}

The proposed platform aims to solve the current problems mentioned above by providing a blockchain-based platform that allows landlords and tenants to manage real estate transactions, through an application that interacts with a Hyperledger Fabric blockchain. Parties can create, sign, manage and delete rental agreements over rental units through a smart contract. The platform is composed by two main components, a front-end enabling interaction between participants, and a complex back-end for processing data. The back-end interacts with multiple components, such as, a Hyperledger Fabric blockchain, a dedicated database that stores participants' personal information, and some external entities such as an external authorization provider and wallets for transacting crypto payments.

\subsection{Architecture}

\begin{figure}[h!]
    \centering
    \includegraphics[width=0.42\textwidth]{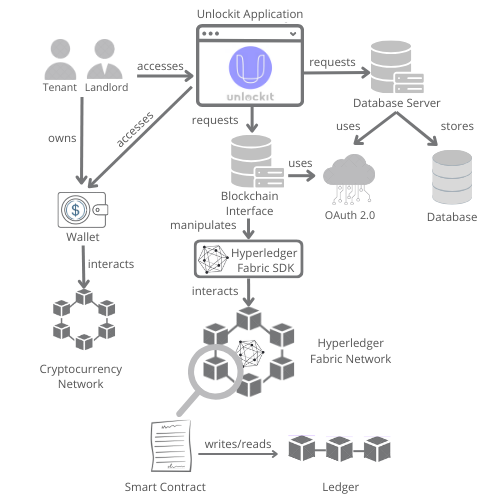}
    \caption{Platform architecture}
    \label{tab:solutionArchitecture}
\end{figure}

The platform high-level architecture represented in figure \ref{tab:solutionArchitecture} illustrates the interaction between participants. Upper in the figure we observe that both landlords and tenants can access the Unlockit application to manage their real estate transactions. The application interacts with three different components that are interconnected to fulfill each other's needs. The application sends requests to two back-end servers and these dispatch them to other components. \\

\subsubsection{\textit{Database Server}}
\label{subsec:javaServer}

The rightmost server in the figure \ref{tab:solutionArchitecture} represents a server interacting with a protocol and a component, namely, OAuth 2.0 and a database. OAuth 2.0 \cite{rfc6749} is an authorization protocol, provided by a trusted third-party entity, that authorizes users to access the application. This protocol allows users to grant Unlockit's application limited access to their third-party provider account without sharing their credentials. It provides a secure way for users to log into Unlockit's application using their third-party provider account information. This process provides the user with an authorization code and an access token. The token is added to each server request to guarantee the user's continuous authorization and authenticity.

Regarding the database, it was created to store and maintain data within three interconnected tables. These tables, namely, ``User'', ``Advertise'' and ``PropertyPhoto'' are linked to each other and to the records stored on the blockchain. The first table stores and manipulates information related to user registration, personal information, and attributes required for the application's business logic. The second table stores and manages information related to an advertisement's publication. And finally, the last table stores only images associated with a property, published in an advertisement. We will revisit these tables later in section \ref{subsec:dataStorageArchitecture}.\\

\subsubsection{\textit{Blockchain Interface}}
\label{subsec:nodeServer}

Looking back to the figure \ref{tab:solutionArchitecture}, we can observe the other server right below the application, referred to as the blockchain interface. It is responsible for interacting with the Hyperledger Fabric blockchain through the SDK. The SDK is the software development kit that provides APIs for our platform to submit transactions, query the ledger, and listen for events emitted by smart contracts on the network.
The blockchain interface server receives requests from the application, processes them and then submits the requests through the SDK API to the Hyperledger Fabric network. Similarly to the database server, this server, interacts with the OAuth 2.0 authorization protocol to authorize users. All requests from the application require this authorization, which is the first security barrier between the application and the blockchain. 

The server provides the application with a simple API containing four endpoints: ``Signup'', ``Login'', ``Evaluate'' and ``Submit''. Table \ref{tab:nodeServerEndpoints} illustrates the server's API with a short description of each one.

\begin{table}[h!]
    \centering
    \caption{Blockchain interface server endpoints}
    \label{tab:nodeServerEndpoints}
    \setlength{\tabcolsep}{10pt}
    \renewcommand{\arraystretch}{1.8}
    \begin{tabular}{|l|p{6.3cm}|}
         \hline
         \textbf{Endpoint}  &  \textbf{Description}\\
         \hline
         signup &  Authorizes user, registers on blockchain, creates credentials and identities, stores in wallets and creates encrypted user id \\
         \cline{1-2}
         login &  Authorizes user and verifies encrypted user id \\
         \cline{1-2}
         evaluate &  Authorizes user and executes chaincode read-only functions \\
         \cline{1-2}
         submit  & Authorizes user, verifies encrypted user id and executes chaincode write-only functions  \\
         \hline
    \end{tabular}
\end{table}

The first endpoint is only used when a new user is registered in the application for the first time. It starts by authorizing and authenticating the user using the OAuth 2.0 protocol. It then registers the user on the blockchain and assigns him his credentials, which are represented by a pair of public and private keys, where the public key is represented by a digital certificate. These credentials are used to generate two identities, a public and a private one. The public identity only holds the digital certificate that allows to retrieve the public key. The private identity stores both the digital certificate and the private key. Both identities are extremely important for the security of users and the blockchain. They are stored in two separate file system wallets, A public accessible to the blockchain, and a private accessible to the user only.

After this, the user's public key is used to encrypt the user id and the result is stored on the blockchain. This step provides the second barrier of protection, ensuring that every transaction made by the user to change the state of the blockchain is validated. This validation is necessary whenever a request is made to the blockchain to change its state, and consists of verifying whether the user encrypted id on the blockchain, when decrypted using the user private key, matches the original user id. If and only if this is verified, the submission is sent.

\begin{figure}[h!]
    \centering
    \includegraphics[width=0.34\textwidth]{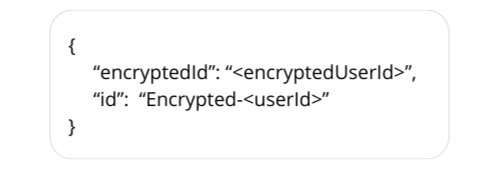}
    \caption{Record of an encrypted id on the blockchain}
    \label{tab:encrypted}
\end{figure}

Finally, a request is sent to the smart contract/chaincode to execute the function ``CreateEncryptedId'' and persist the data in the blockchain world state. This function registers the encrypted Id in the format shown in figure \ref{tab:encrypted}. If everything goes as expected, a response is sent to the application and the user may start using it. Further on, we will explain more about the smart contract/chaincode.

The second endpoint, ``login'', is used whenever it is necessary to log in to the application. The blockchain network must always be active to allow users to log in to the application. The endpoint begins by authorizing and authenticating the user using the OAuth 2.0 protocol. It then submits a read request to the smart contract/chaincode function ``ReadAsset'' to read the user encryptedId record, previously created during the ``signup'' endpoint. Then, checks whether the user's private wallet exists and extracts the private key. The private key is then applied to the encrypted user id, resulting in the decrypted user id.  It then checks whether this decrypted user id equals the original user id. If so, the user's identity is verified and the user is logged into the application. 

The third endpoint, ``evaluate'', does not require changing the network's world state. It only executes read functions. It starts by authorizing the user using the OAuth 2.0 protocol. If successful, runs one of the read functions provided by the smart contract/chaincode. The function to be executed depends on the application's needs when requesting the blockchain interface server.

The fourth and final ``submit'' endpoint is used whenever a user wants to make a change to the world state of the blockchain network. It authorizes the user using the OAuth 2.0 protocol, and then, validates him by verifying its encrypted id record, just like in the process performed during the login endpoint. If validated, the user is able to submit a transaction. In other words, the server is able to call a smart contract/chaincode write-function. And once again, the function to be executed depends on the application's needs when requesting the blockchain interface server.\\

\subsubsection{\textit{Hyperledger Fabric Blockchain}}
\label{subsec:hyperledgerFabricBlockchain}

Taking another look at the figure \ref{tab:solutionArchitecture}, we find the Hyperledger Fabric network representation below its SDK. We implemented and deployed a Hyperledger Fabric blockchain on a Kubernetes cluster. 
A Kubernetes cluster is a set of interconnected physical or virtual machines running Kubernetes, an open source container orchestration platform. We chose to implement the Hyperledger Fabric blockchain on a Kubernetes cluster to simplify the management and scalability of its components, such as peers, orderers and CA's, guaranteeing the network's adaptability to changing workloads.

For simulating a Kubernetes cluster without the overhead of running it on separate virtual machines or cloud instances, we used the KinD tool, short for Kubernetes in Docker. It's a lightweight tool designed for local development, testing, and debugging. It allows to create, run and manage a kubernetes cluster inside a Docker container. In our setup, we employed KinD to establish a Kubernetes cluster. The cluster enables the deployment of Hyperledger Fabric network elements, effectively simulating a real network environment locally.

\begin{figure}[h!]
    \centering
    \includegraphics[width=0.4\textwidth]{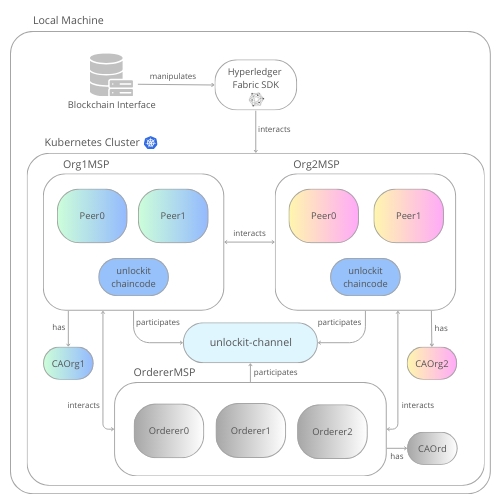}
    \caption{Hyperledger Fabric network on kubernetes}
    \label{tab:hyperledgerFabricSolutionNetwork}
\end{figure}

Figure \ref{tab:hyperledgerFabricSolutionNetwork} illustrates a local environment of a machine running our platform. Inside the machine we can observe the blockchain interface server manipulating the Fabric network SDK to interact with the network hosted on a KinD cluster. In this diagram we omitted the remaining components of the architecture depicted in figure \ref{tab:solutionArchitecture}, to focus our attention on the Hyperledger Fabric network.

The Hyperledger Fabric network runs inside a KinD cluster and contains three organizations, Org1, Org2, Org3. Each one has its own Membership Service Provider and Certification Authority, represented in the figure \ref{tab:hyperledgerFabricSolutionNetwork} by Org1MSP, Org2MSP, OrdererMSP, CAOrg1, CAOrg2 and CAOrd respectively. The CAs manage user identities, issue certificates and handle authentication processes within the organization. All organizations participate in the same channel, which consequently implies that all peers within those organizations participate in the same channel as well. Both organizations, Org1 and Org2, have the same number of peers, Peer0 and Peer1, and the same chaincode deployed on them. Each peer maintain a copy of the ledger and a copy of the chaincode associated with the channel. The organization marked in grey, represents the organization in charge of the network's consensus. It is composed by three ordering peers who establish the consensus of the transactions, validating and maintaining the network.

In our platform, when the application submits a transaction to the Fabric network, the request is sent to the blockchain interface server, which uses the SDK to submit the request to either organization 1 or 2. The selected one, sends the transaction proposal to its two peers for approval. They simulate it and provide an approval signature. Once approved, the SDK sends the approved transaction to the ordering organization's to be ordered by its ordering peers. At this stage, the ordering peers, group the transactions into blocks, reach a consensus on the order of the transactions and then create blocks containing the endorsed transactions. Afterwards, the blocks are disseminated to the peers of all the participating organizations (Org1, Org2) for validation and commitment to their ledgers.

Our system is specifically designed to allow state changes in the blockchain solely through transaction submissions to Organization 1. Consequently, Organization 1 holds the read and write permissions, while Organization 2 is limited to read-only access. Essentially, Organization 2 can only execute query transactions within the network. This deliberate design ensures that the Organization 1, (Unlockit organization), retains exclusive authority over ledger modifications, while the remaining organizations would exclusively act as network auditors.\\

\subsubsection{\textit{Smart Contract}}
\label{subsec:hyperledgerFabricSmartContract}

As mentioned in the previous section, our Hyperledger Fabric network contains only one smart contract in charge of interacting with the assets stored on the blockchain network. Assets are digital representations of real-world objects such as a real estate property. They are represented as a collection of key-value pairs in binary and/or JSON format. Our platform is designed to contain six assets, namely, ``PropertyAsset'', ``ContractAsset'', ``Proposal'', ``RentalInfo'', ``Payment'' and ``EncryptedId'', all registered on the blockchain through different chaincode functions.

The ``PropertyAsset'' record represents a physical property. The ``ContractAsset'' record encapsulates digital rental contract details, including contractual information and digital signatures from both parties. The ``Proposal'' record represents a tenant's proposal for a particular advertisement. The ``RentalInfo'' record provides an analytical analysis of all proposals made for an advertisement. It is only created when a rental contract is confirmed and records two metrics, the number of proposals made to an advertisement by all potential tenants, and the proposal with the highest value. The ``Payment'' record, registers the payments linked to a rental contract. Finally, the ``EncryptedId'' record, as explained in the section \ref{subsec:nodeServer}, serves to verify users' identities before letting them submit transactions to change the state of the blockchain. 

Lets now examine how the chaincode manipulates the previous assets. Hyperledger Fabric chaincode works similarly to a common API, offering CRUD (Create, Read, Update, Delete) operations. However, it distinguishes itself due to the fact that, data management is specifically targeted towards the blockchain.

Our chaincode offers an API containing several public and private functions targeting different records. The private ones are only accessible from within the chaincode, meaning that they can be called by other chaincode functions, whereas, the public ones can be called by the Fabric SDK through an application request.

Every public function altering the blockchain state validates the user id. When the blockchain interface server invokes any of these functions, sends a request, with the user id as the first argument, previously verified by the server. During execution, this id is compared with another id, also provided as an argument. The function proceeds only if the ids match. To make it clearer, we will demonstrate with two functions.

The ``CreatePropertyAsset'' function receives the user id as the first parameter and the other parameters relate to the property being created. One of them is the ``landlordId'', which must match the user id of the user submitting the request. The function only proceeds its execution if the user id is equal to the ``landlordId''. Let us now take as an example the ``DeletePropertyAsset'' function which receives two parameters, the user id and the property id. Note that, only the owner of the property must be able to execute this function. To verify this, we use a private chaincode reading function that retrieves the JSON of the property's asset. From here, we obtain the ``landlordId'', since the property asset contains this information. Finally, we compare the two ids and check whether they are the same. 

Both of these examples are a sample of what is done similarly in other public chaincode functions. By doing so, we guarantee that all landlords are able to create properties, but only the landlord who owns a given property can delete it. We guarantee that all landlords are able to create rental contracts, but only the landlord of a given contract can delete and update it. We guarantee that all tenants are able to submit proposals, but only the tenant who submitted a given proposal can delete it. Finally, we guarantee that all landlords are able to create payments, but only the landlord and tenant of a given payment can update and delete it. 

There are also other constraints that apply to specific public functions within the chaincode. For instance, a property, a contract, or a payment record, cannot be deleted if there is an active rental contract. Moreover, a user cannot delete his data from the platform if he is involved in an ongoing rental contract, or if he has pending payments to receive or send.\\

\subsubsection{\textit{Data Storage Architecture}}
\label{subsec:dataStorageArchitecture}

\begin{figure}[h!]
    \centering
    \includegraphics[width=\columnwidth]{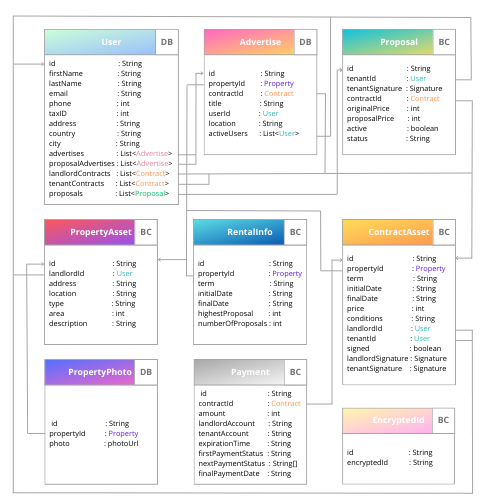}
    \caption{The data storage architecture}
    \label{tab:dataStorageArchitectureIlustration}
\end{figure}

Figure \ref{tab:dataStorageArchitectureIlustration} shows a compact diagram of how data is managed and stored on the different components of our platform.

 It offers a detailed perspective on the interconnections between database tables and their relationships with the blockchain records. Each square is labeled with the table or record name, the storage type, and multiple attributes. The storage type is an acronym represented on the square's top right corner indicating ``DB'' for Database and ``BC'' for Blockchain. The relationships between squares are represented by arrows, whose tips indicate ownership or reference.\\

\subsubsection{\textit{Rental Payments}}
\label{subsec:payments}

Traditional payment methods have long been a fundamental aspect of financial transactions. In recent times, there has been a notable shift towards cryptocurrency payments, offering a decentralized and secure alternative. Cryptocurrency payments, particularly in stablecoins like USDC and USDT, have gained substantial traction due to their inherent stability tied to real-world assets, addressing the price volatility often associated with cryptocurrencies. 

In the context of our platform, we focus on stablecoin payments on a cryptocurrency network. In the figure \ref{tab:solutionArchitecture}, each user of the application has access to its own wallet, serving as a storage space for their stablecoins. The wallet's primary function is to engage with the cryptocurrency network by submiting and receiving transactions. Our platform enables users to access their wallets securely through the application, since it integrates various wallet types. 

To perform a transaction its necessary to know two pieces of information about the receiver. The amount to be sent and the address to which we want to send the amount. Tenants make rental payments by sending funds from their wallets to the landlord's cryptocurrency network address. We chose to execute payments on a cryptocurrency network due to its open nature, promoting a democratic financial ecosystem. 

\begin{figure}[h!]
    \centering
    \includegraphics[width=0.25\textwidth]{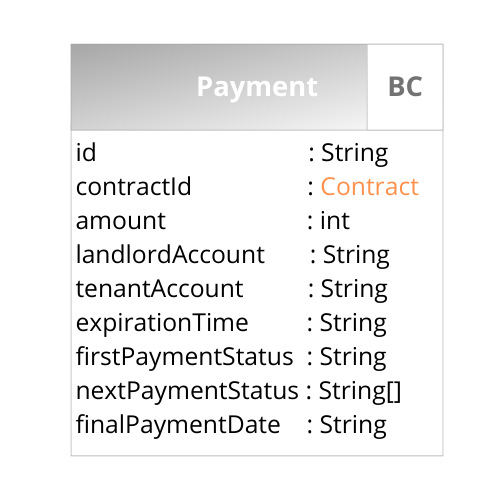}
    \caption{Payment record}
    \label{tab:paymentRecord}
\end{figure}

Our platform maintains a payment record on the Hyperledger Fabric blockchain associated to a rental contract. Figure \ref{tab:paymentRecord} shows an expanded version of the payment record extracted from figure \ref{tab:dataStorageArchitectureIlustration}.

Each payment record contains: a reference to the contract id linked to it, the amount to be paid, the sender and recipient accounts represented by their cryptocurrency network address, the first payment's expiration time, the first and subsequent payment's transaction status, if applicable, and the payment's final date. 

There are two types of rental contracts, long and short term. Depending on the rental type, the payment will be executed monthly or in one single sum. If it is a short-term contract, it will last less than a month, and therefore the payment will only be made once. On the other hand, if it is a long-term contract, it will last longer than a month and the payment will be made on several occasions. In this case, after the initial payment, the smart contract will process all subsequent payments on the first day of each month. 

The payment record is adaptable to different types of contracts. It maintains a status for each rental payment. The status represents whether or not a given transaction has already been confirmed on the cryptocurrency network. All payment records hold a status for the first payment. However, only long-term contracts require their payment record to hold the status of the following rental payments. 

As an example, a short term contract, lasting 25 days, has a payment record holding a single reference to the status of the first payment transaction, whereas a long term contract lasting, say, 3 months, has a payment record holding reference to the status of the first and the two subsequent payment transactions. The first payment is coupled with an expiration time defined by the landlord on the platform, which represents the time limit within which, the tenant must make the payment. Finally, the last attribute is used to identify the end date of a contract's last payment.

The first payment is always generated by the landlord through the application. The application sends the request to the blockchain interface server, which then invokes the chaincode function to create a payment. However, in order to generate the payment for the following monthly rents, it is necessary to automate the process, avoiding manual user interaction. 

Our platform uses an external scheduler service, a cron job in kubernetes, to invoke the ``processMonthlyPayments'' chaincode function at midnight in the first day of each month, processing all pending payments for all rental contracts.\\

\subsubsection{Rental Process}
\label{subsec:rentalProcess}

\begin{figure}[h!]
    \centering
    \includegraphics[width=0.44\textwidth]{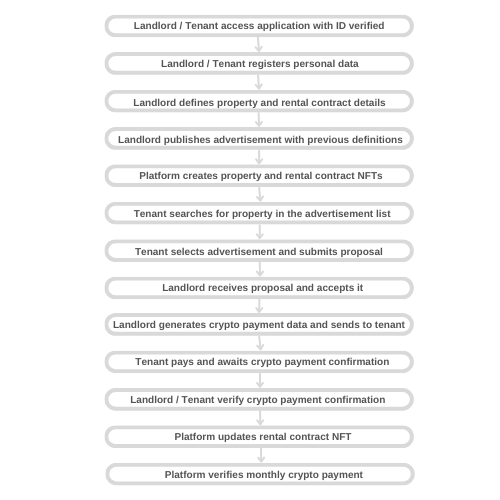}
    \caption{The rental process}
    \label{tab:process}
\end{figure}

Figure \ref{tab:process} illustrates the steps required to establish a rental contract. Users, whether tenants or landlords, log into the application through a third-party provider's authorization and authentication protocol, which assigns them with their third-party provider's user unique id.

Once authenticated, users can access the application to consult the advertisements list. However, they are only able to create advertisements and submit rental proposals after registering their personal information in the application.

After registering their personal data, landlords may start creating advertisements. Creating an advertisement requires defining the property and rental contract details. Ideally, the property details should be unique, to ensure a unique representation of the property on the blockchain. Although not implemented in our platform, a property should be verified before being published by checking, for instance, its title deed or tax registration. The platform permits advertising different types of properties, ranging from entire houses to different apartment typologies and even single rooms. It supports fixed-term, month-to-month, short-term and room rental types. Once defined all the advertisement details, the landlord publishes it on the platform. Three records are created during the publication, namely, advertisement, property and rental contract. The advertisement record, is stored in the database and holds two references, one for the property and the other for the rental contract. The property and the rental contract records are stored on the Fabric blockchain network, and are represented as NFTs. 

The rental contract NFT is assigned with the landlord's digital signature. The contract signature is materialized through a ciphered message representing the details of the rental contract. The message is a concatenated string containing a few contract details such as the, contract id, property id, term, initial, final dates, and conditions. The digital signature is therefore generated by applying the SHA256 hash, chosen for its security, to the concatenated string, followed by the user’s private key. The landlord's digital signature applied to a rental contract guarantees the non-repudiation of the landlord who signed it. A tenant wishing to verify the landlord's digital signature is allowed to do it, by applying the landlord's public key to the landlord's digital signature.

Once an advertisement is active, a potential tenant consults the advertisements list and submits a digitally signed proposal to the advertisement. Since the proposal record contains a reference to the contract record, the tenant produces a digital signature for the same concatenated string as the landlord did when creating the contract. The proposal represents the tenant's acceptance of the contract terms set by the landlord, together with the adjustable rental amount that the tenant wishes to submit.

The landlord is notified in the application about a potential tenant's interest and checks the proposal submitted. Note that, the landlord may have several proposals for the same advertisement. At this point, the landlord can opt to either reject or accept the proposal. If rejected, the potential tenant will be notified in the application and may resubmit another proposal for the same advertisement. If accepted, the advertisement will be temporarily unavailable to other users until payment is confirmed or rejected. Once the proposal has been accepted, the landlord inserts his payment details including a configurable payment time limit and sends it to the tenant. If the payment transaction is not registered on the cryptocurrency network within the configured time limit, the proposal is rejected and the advertisement is made available to all users again. As for the payment details, the amount will be automatically displayed on the platform according to the tenant's proposal amount. The landlord simply has to select the stablecoin (USDC, USDT) he prefers to be paid in, and enter his cryptocurrency network account public key.

The payment details together with the payment deadline are sent to the tenant's email address. After receiving the details, the tenant connects his wallet through the application, enters the payment details and submits the transaction. The wallet can be connected by opening its browser extension or by scanning a QR code to open it on the tenant's mobile device. Once the transaction is sent to the cryptocurrency network, the proposal can no longer be rejected, unless the transaction expires, which does not usually happen.

The time for confirming a transaction on a cryptocurrency network is variable, therefore, we can not tell when a transaction will be effectively confirmed. Depending on the cryptocurrency network implemented, a transaction may take seconds, minutes or hours to be confirmed. Therefore, both parties have to be constantly verifying it until one party actually confirms it. As soon as one of the parties verifies the transaction confirmation, the platform updates the rental contract's NFT, by adding the new tenant id along with the tenant digital signature obtained from the proposal previously submitted by the tenant.

Depending on the rental type, long or short term, the payment will be executed monthly or in one single sum, as explained in the section \ref{subsec:payments}. If it is a short-term contract the payment will only be made once, while, if it is a long-term contract, it will be made on several occasions. In this case, after the initial payment, the smart contract will process all subsequent payments on the first day of each month. All subsequent payment details are generated using the same information as the first one, including the landlord's cryptocurrency network address and the exact rental amount. However, unlike the first, the subsequent payments do not require a payment time limit.

\subsection{Implementation}
\label{section:implementation}

The described architecture was implemented using different technologies. We used 
the React framework to implement the application front-end. For the back-end, we utilized OAuth 2.0 protocol provided by Google. The database server was written in java and interacts with a database provided by Mongodb. The blockchain interface server was written in JavaScript/Typescript and runs on Node runtime environment, which interacts with the Hyperledger Fabric network through an SDK.

Regarding payments, the application integrates a Web3 wallet through a browser extension or a mobile application for processing payments using Stablecoins in Solana, a high-performance blockchain platform. Solana was chosen due to its fast transaction confirmation times, processing thousands of transactions per second, making it an efficient choice for real-time transactions. Additionally, Solana's low gas fees enhance the cost-effectiveness of cryptocurrency payments, making it an attractive option for our application and for testing purposes.

\section{Platform Evaluation}

We evaluate the platform in terms of to its adherence to the European Union's General Data Protection Regulation and performance benchmarks. Our goal, is to provide information on the applicability of the platform in the real world.

\subsection{GDPR Compliance}
\label{section:gdprCompliance}

Our platform is designed to provide real estate services to European citizens. Consequently, it manipulates customer sensitive data. Therefore, it must comply with the GDPR, fulfilling the seven principles of protection and accountability and the eight user privacy rights. Let us start by examining the platform compliance with the seven data protection principles.\\

\begin{enumerate}

\item \textit{Legality, Fairness, and Transparency -}
Data processing must be legal, fair, and transparent to the individuals concerned. Most of the processed data, such as contracts, properties, proposals and payments are stored on the blockchain. The blockchain ensures the principle of transparency and fairness, as data processing can be viewed by an authorized individual. With regard to legality, personal data is collected only if the user wishes to continue on the platform to begin renting or listing properties.

\item \textit{Purpose Limitation -}
Data should only be processed for legitimate, explicit purposes communicated to the data subjects during collection. Personal data collected by the platform is only acquired to identify a user and its collection is explicitly stated on the platform.

\item \textit{Data Minimization -}
Collect and process only the necessary data essential for the specified purposes. Personal data is collected for user identification purposes only and data relating to property, rental contracts and payments are collected solely for the purpose of establishing rental contracts.

\item \textit{Accuracy -}
Personal data must be kept accurate and kept up-to-date. The platform keeps up-to-date personal data in the database, and the latest blockchain records in the world state.

\item \textit{Storage Restriction -}
Store personally identifiable data only for the necessary duration required for the stated purpose. The user's personal data is kept until the end date of the last active rental contract or the last pending payment. If the user has no active rental contracts, or pending payments, his personal data is kept until he quits the platform.

\item \textit{Integrity and Confidentiality -}
Processing must guarantee appropriate integrity, and confidentiality, such as through encryption methods. The platform keeps personal data encrypted and secure in the database. Property, rental contracts, proposals and payment records are secured in the blockchain. The integrity of blockchain transactions is ensured by constantly validating users' encrypted IDs. The database tables are encrypted by an encryption key stored within the database, which is protected by a master key stored in a keyfile on the MongoDB server.

\item \textit{Accountability -}
The data controller is accountable for demonstrating compliance with these principles as per GDPR regulations. Blockchain records are transparent and therefore available for auditing at any time. Users' personal data, on the other hand, can be requested for auditing from Unlockit whenever necessary.
 
\end{enumerate}

Next, we evaluate compliance with the eight GDPR privacy rights, which empower individuals with increased control over the data they share with Unlockit.\\

\begin{enumerate}

\item \textit{The right to be informed -} 
Individuals have the right to know how their data will be used. The platform collects personal data with users' consent. Users will only be able to access its services, once they authorize the collection of their personal data.

\item \textit{The right of access -}
Individuals can request access to their personal data that an organization holds. A user may access, verify and update his personal data at any time on his profile page.

\item \textit{The right to rectification -}
Individuals can request corrections to inaccurate or incomplete data. A user can rectify and update his personal data on his profile page at any time.

\item \textit{The right to erasure (The right to be forgotten) -}
Individuals can request the deletion of their personal data. Users are identified by a unique id, linking their personal data in the database with the associated blockchain records. This right allows a user to delete all his data from the platform whenever he wishes. However, there are some restrictions imposed by the platform. A user can only delete his information if no rental contracts are active and no payments are pending. Should this not be the case, the user must wait for the last active contract to end or complete any missing payments. If these conditions are met and the user wishes to delete his data, he may do so on his profile page. By deleting his account, his personal data is erased from the database, and the blockchain records associated with the user id, such as property, rental contracts, payments, and proposals, are also deleted from the world state. However, they are always available for auditing in the transaction log. 

\item \textit{The right to restrict processing -}
Individuals can limit the way an organization uses their data. By entering their data, users grant the platform permission to process it for identification purposes only. Should the platform use the data for another purpose, users would eventually, be able to configure their usage. At the moment, this right is not applicable.

\item \textit{The right to data portability -}
Individuals can obtain and reuse their data for their own purposes. Users may access their personal data from the profile page anytime, allowing them to use profile data for purposes beyond the platform.

\item \textit{The right to object -}
Individuals can object to the processing of their data in certain circumstances. This right does not apply to our current platform, since users authorize the handling of their data solely for the purpose of user identification. Therefore, there are no other circumstances to object to.

\item \textit{Rights in relation to automated decision making and profiling -}   
Individuals have safeguards against the risk of a potentially damaging decision being made without human intervention. Since our application does not have automated decision-making and profiling mechanisms, this right does not apply.

\end{enumerate}

In summary, the preceding assessment ensures that our platform correctly processes users' personal data, guaranteeing full compliance with the EU GDPR's data protection principles and user privacy rights. 

\subsection{Platform Performance}
\label{section:platformPerformance}

We evaluate the platform's API through performance tests using key metrics such as response time, throughput and error rate. To conduct these measurements, we used Postman performance testing tool, to simulate real-world traffic. The tests were performed on a local execution environment running the platform, consisting of a virtual machine running Linux Mint Mate 21.2, 64-bit OS with 3 virtual processors, 6.4 GB of RAM and 150 GB of storage. The virtual machine is hosted on a Linux Mint Cinnamon 20, 64-bit OS, equipped with a dual-core Intel processor operating at a base clock speed of 1.60 gigahertz, with 7.5 GB of RAM and 480 GB of storage.

The platform has a collection of twenty-five public functions, available for testing, distributed across three APIs: the smart contract, the database server and the blockchain interface server. However, although indispensable, presenting independent tests for all of them, would overload this document. Therefore, we decided to test the platform's two most significant processes, which involve accessing various public and private functions. They are, the process of publishing a property and the process of submitting a proposal.

As explained in section \ref{tab:solutionArchitecture}, the process of creating an advertising involves authorizing the user in the application, validating his identity on the blockchain, creating the property and rental contract records on the blockchain and updating all the database tables related to the landlord that publishes the advertisement. As for submitting a proposal, it also involves authorizing the user in the application, validating his identity on the blockchain, creating a proposal record on the blockchain and updating the advertise and user database tables related to the tenant that submits the proposal.

We measure the average response time, throughput and error rate for each of these processes. The tests are executed with an increasing number of virtual users, ranging from 1, 10, 20, 40, 60, 80 up to a maximum of 100 virtual users. For each of these sets of virtual users, which send parallel requests to the platform, we measured 10 times, each lasting 1 minute, and calculated the metrics average, producing two graphs.

\begin{figure}[h!]
    \centering
    \includegraphics[width=0.5\textwidth]{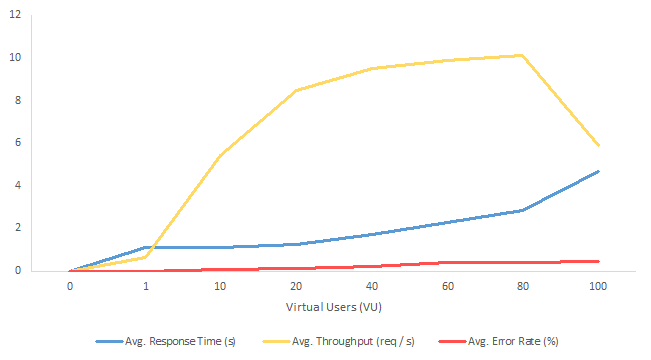}
    \caption{Publish advertisement - performance}
    \label{tab:publishAdvertisementProcessPerformance}
\end{figure}

Each graph plots one variable against the number of virtual users, where the x-axis indicates the number of virtual users (VU) and the y-axis shows performance metrics, such as, response time (blue line), throughput (golden line) and the error rate (red line). The first graph, figure \ref{tab:publishAdvertisementProcessPerformance}, measure the average response time and throughput that it takes to publish an advertisement. This involves evaluating some of the platform functions in the following order: ``CreateContractAsset'' (smart contract), ``CreatePropertyAsset'' (smart contract), ``registerPropertyPhoto'' (database), ``getUserById'' (database), and ``registerAdvertise'' (database). Table \ref{tab:publishAdvertisementProcessPerformanceSummary} summarizes, for each metric, the average results obtained from the performance tests conducted for the process of publishing an advertisement.

The blue line describes the average response time, that peaks, when VU count is at its peak measuring approximately 4.6 s. The golden line describes the average throughput, that increases as the number of VUs rises, reaching the peak of 10 req/s at around 80 VUs. After this point, the average throughput starts to decrease as the platform becomes overloaded and is no longer able to handle the load effectively. The red line describes the average error rate, that remains low, increasing as the number of virtual users grows, reaching a maximum error rate of approximately 0.45\% when VUs peak at 100.

\vspace{-2pt}

\begin{table}[h!]
    \centering
    \caption{Publish advertisement - performance summary}
    \label{tab:publishAdvertisementProcessPerformanceSummary}
    \setlength{\tabcolsep}{3.3pt}
    \renewcommand{\arraystretch}{1.6}
    \begin{tabular}{|l|l|l|l|}
         \hline
         \textbf{Avg. Req. Sent} & \textbf{Avg. Throughput} & \textbf{Avg. Resp. Time} & \textbf{Avg. Error Rate} \\
         \hline
        664  & 7.13 req/s &  2.14 ms & 0.24 \% \\
         \cline{1-4}
         \hline
    \end{tabular}
\end{table}

\vspace{-18pt}

\begin{figure}[h!]
    \centering
    \includegraphics[width=0.5\textwidth]{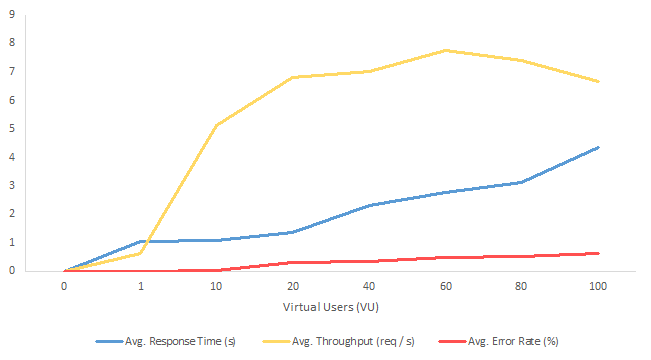}
    \caption{Submit proposal - performance}
    \vspace{-10pt}
    \label{tab:submitProposalProcessPerformance}
\end{figure}

The second graph, figure \ref{tab:submitProposalProcessPerformance}, measures the average response time, throughput and error rate that it takes to submit a proposal. This involves evaluating some of the platform functions in the following order: ``CreateProposal'' (smart contract), ``updateUserById'' (database), and ``updateAdvertiseById'' (database). Table \ref{tab:submitProposalProcessPerformanceSummary} summarizes, for each metric, the average results obtained from the performance tests conducted for the process of submitting a proposal.

\begin{table}[h!]
    \centering
    \caption{Submit proposal - performance summary}
    \label{tab:submitProposalProcessPerformanceSummary}
    \setlength{\tabcolsep}{3.3pt}
    \renewcommand{\arraystretch}{1.6}
    \begin{tabular}{|l|l|l|l|}
         \hline
         \textbf{Avg. Req. Sent} & \textbf{Avg. Throughput} & \textbf{Avg. Resp. Time} & \textbf{Avg. Error Rate} \\
         \hline
        569  & 5.90 req/s &  2.29 ms & 0.33 \% \\
         \cline{1-4}
         \hline
    \end{tabular}
\end{table}

\vspace{-2pt}

The blue line, shows the average response time increasing as the number of VUs increases. The highest average response time observed is approximately 4.3 s. This suggests that the platform is becoming saturated as more users access it, indicating a potential correlation between load and response time which can be a concern for scalability. The golden line describes the average throughput increasing linearly as the number of VUs increases until it reaches a peak at around 60 VUs, registering approximately 7.7 req/s. This indicates that beyond 60 VUs, the average throughput starts to decrease, suggesting that the platform reached its maximum capacity. The red line describes the average error rate, that increases gradually as the number of VUs rises, reaching a maximum error rate of approximately 0.61\% when VUs peak at 100.

Both process tests show a significant error rate, which indicates that the platform might need optimization or resources enhancement to reduce this error rate. According to our measurements, only the functions that submit transactions to the smart contract experience failures. The most common and unique error is the, ``ECONNRESET'', which stands for ``Connection Reset by Peer''. It occurs when a TCP connection between two machines is unexpectedly terminated by the peer, which can happen when there is network congestion due to high resource utilization, causing the connection to reset. Considering this evidence, in a future iteration we would recommend increasing the number of peers in organization 1 and the number of ordering peers in organization 3, to handle a higher volume of requests.

In this evaluation we demonstrated the platform's compliance with the European data protection regulation (GDPR) and tested the platform's performance when exposed to a high number of parallel requests. We concluded that refining the Hyperledger Fabric blockchain network could enhance the performance of the analyzed processes, making the platform more suitable for real-world deployment.

\section{Conclusion}

This paper presented a platform design conceived for Unlockit, a DLT company operating in the real estate industry. We proposed a blockchain-based residential smart rental platform, designed to solve the existing residential rental problems through the latest smart contract technology supported by a permissioned blockchain, guaranteeing the integrity, security and transparency of real estate rental transactions. We presented the key concepts needed to understand the platform, followed by a presentation of its architecture and implementation and concluded with its evaluation, assessing the platform compliance with the GDPR and performance.

\section{Acknowledgements}

This paper was developed within the scope of the project nr.\ 51 ``BLOCKCHAIN.PT - Agenda Descentralizar Portugal com Blockchain'', financed by European Funds, namely ``Recovery and Resilience Plan - Component 5: Agendas Mobilizadoras para a Inovação Empresarial'', included in the NextGenerationEU funding program.



\end{document}